\documentclass{emulateapj}
 \submitted{{\it Accepted to the Astrophysical Journal on November 11, 2014}}

\begin{document}
\title{A HERO'S LITTLE HORSE: DISCOVERY OF A DISSOLVING STAR CLUSTER IN PEGASUS}
\author{Dongwon Kim\altaffilmark{1} and Helmut Jerjen\altaffilmark{1}}
\affil{Research School of Astronomy and Astrophysics, The Australian National University, Mt Stromlo Observatory, via Cotter Rd, 
Weston, ACT 2611, Australia}

\email{dongwon.kim, helmut.jerjen@anu.edu.au}

\begin{abstract}
We report the discovery of an ultra-faint stellar system in the constellation of Pegasus. This concentration of stars was 
detected by applying our overdensity detection algorithm to the SDSS-DR10 and  confirmed with deeper photometry 
from the Dark Energy Camera at the 4-m Blanco telescope. The best-fitting model isochrone indicates that this stellar system, Kim\,1, features an old (12\,Gyr) and metal-poor ([Fe/H]$\sim-1.7$) stellar population at a heliocentric distance of $19.8\pm0.9$\,kpc. 
We measure a half-light radius of $6.9\pm0.6$\,pc using a Plummer profile.  The small physical size and the extremely 
low luminosity are comparable to the faintest known  star clusters Segue\,3, Koposov\,1 \& 2, and Mu\~noz 1. However, 
Kim\,1 exhibits a lower star concentration and is lacking a well defined center. It also has an unusually high ellipticity
and irregular outer isophotes, which suggests that we are seeing an intermediate mass star cluster being stripped 
by the Galactic tidal field. An extended search for evidence of an associated stellar stream within the 3 sqr deg DECam 
field remains inconclusive. The finding of Kim\,1  is consistent with current overdensity detection limits and supports the hypothesis 
that there are still a substantial number of extreme low luminosity star clusters undetected in the wider Milky Way halo.
\end{abstract}


\section{Introduction}
Modern imaging surveys like the Sloan Digital Sky Survey~\citep{York2000,Ahn2014} have significantly contributed to the discoveries of new stellar objects in the Milky Way halo including satellite galaxies~\citep[e.g][]{Willman2005,Belokurov2006,Irwin2007,Walsh2007} and star clusters~\citep{Koposov2007,Belokurov2010,Munoz2012,Balbinot2013,Belokurov2014,Laevens2014}. The new satellite galaxies are characterised by low luminosities ($ -8 \lesssim M_{V} \lesssim -1.5 $)~\citep{Martin2008} and low metallicities down to [Fe/H]$<-3$~\citep{Kirby2008,Norris2010,Simon2011,Koch2014} . The emergence of this new class of dwarf galaxies now questions previous ideas about the low mass limit of galaxy formation. The other new MW halo objects are star clusters with extremely low luminosities ($-2.5 \lesssim M_{V} \lesssim 0$) and small half-light radii ($ < 10$ pc). Globular clusters of such extreme nature are thought to be suffering stellar mass loss via dynamical processes such as tidal disruption or evaporation~\citep{Gnedin1997,Rosenberg1998,Koposov2007}, and there is growing evidence based on observations to support this hypothesis ~\citep{Carraro2007,Carraro2009,Niederste2010,Fadely2011}. The discoveries of new ultra-faint star clusters in the Galactic halo will provide valuable resources for studies of their evolutionary phases as well as the formation history of the Galactic halo.

In this paper we present the discovery of a new ultra-faint stellar system named Kim\,1, in the constellation of Pegasus found in SDSS Data Release 10 and confirmed with deep DECam imaging (Sections 2 \& 3). Its total luminosity is measured to be $\sim 0.3$ magnitudes fainter than that of Segue\,3 ($M_{V}=0.0\pm0.8$) known to be probably the faintest star cluster to date~\citep{Fadely2011}. Kim\,1 lies at a heliocentric distance of approximately 20 kpc, is highly elongated ($\epsilon=0.42$) and has a half-light radius of $\sim7$ pc (Section 4). In Section 5 we discuss the possible origin of the stellar overdensity and draw our conclusions.

\section{ Discovery}

The SDSS imaging data are produced in the $ugriz$ photometric bands to a depth of $r\sim22.5$ magnitudes~\citep{York2000}. Data Release 10 (DR10) includes all the previous data releases, covering $14,555 \deg^{2}$ around the north Galactic pole~\citep{Ahn2014}, and is publicly available on the SDSS-III Web site\footnote{http://www.sdss3.org/dr10/}.

 We use an overdensity detection algorithm built upon the method of \citet{Invisibles} to analyse the SDSS-DR10 point source catalogue. The algorithm has been designed to detect stars consistent with an old and metal-poor population, which reflects the typical characteristics of Milky Way satellite galaxies~\citep[see e.g.][]{Kirby2013}. At given heliocentric distance intervals we apply a photometric filter in color-magnitude space employing isochrone masks based on the Dartmouth models~\citep{Dartmouth} and the SDSS photometric uncertainties. We then bin the RA, DEC positions of the selected stars into a $1.0\,^{\circ}\times1.0\,^{\circ}$ array with a  pixel size between $18-25$\,arcsec. This array is then convolved with a Gaussian kernel with a FWHM between $42-59$\,arcsec. We empirically measure the statistical significance of potential overdensities by comparing their signal-to-noise ratios (SNRs) on the array to those of random clustering in the residual Galactic foreground. This process is repeated by shifting the isochrone masks over a range of distance moduli $(m-M)$ from $16$ to $22$ magnitudes, corresponding to the heliocentric distance interval $16<D<250$\,kpc, where the upper limit is the virial radius of the Milky Way.

The detection algorithm successfully recovered all of the recently reported MW galaxy companions in the SDSS footprint as well as other resolved stellar overdensities such as Balbinot 1~\citep{Balbinot2013}. Our measured statistical clustering significances of the MW satellites, including Leo V and Bootes II that were reported as marginal detections by \citet{Invisibles} and \citet{Koposov2008}, were all at least $6.0\,\sigma$ above the Poisson noise of Galactic stars while the significance of a newly found stellar overdensity reached $\sim 6.8\,\sigma$ at its maximum. This object that we chose to call Kim\,1 was found at the location 22$^h$11$^m$41.28$^s$, +07$^d$01$^m$31.8$^s$\,(J2000). It is also worth to mention that we detected a few more stellar concentrations over the 
entire SDSS-DR10 footprint with comparable significance but at larger distance moduli $(m-M)>20$. They will require deeper follow-up imaging observations to reach the main sequence turn-off in the color-magnitude diagram to confirm their  identities.

\section{Follow-up Observations and Data Reduction}
To investigate the nature of Kim\,1, we conducted deep follow-up observations using the Dark Energy Camera (DECam) at the 4-m Blanco Telescope at Cerro Tololo Inter-American Observatory (CTIO) on 17th July 2014. DECam has an array of sixty-two 2k $\times$ 4k CCD detectors with a 2.2 degrees field of view and a pixel scale of 0.27$\arcsec$ (unbinned). We obtained a series of 8$\times$210\,s dithered exposures in $g$ and 5$\times$210\,s  in $r$ band under  photometric conditions. The mean seeing was 1.0$\arcsec$ in both filters. The fully reduced and stacked images were produced by the DECam community pipeline \citep{DECamCP2014}. We used WeightWatcher \citep{WeightWatcher} for weight-map combination and SExtractor \citep{SExtractor} for source detection and photometry. Sources 
were morphologically classified as stellar or non-stellar objects. We cross-matched the extracted point sources with SDSS stars in our field-of-view using the STILTS software \citep{STILTS} with a $1.0\arcsec$ positional tolerance. The photometric calibration was restricted to stars in the magnitude range $19.25<r_{0}<21.25$ mag to stay below the saturation limit of our DECam images and above the $5\sigma$ limit of the SDSS photometry. Finally, all magnitudes of the calibrated sources were extinction-corrected with the \cite{Schlegel1998} extinction maps.

\begin{figure}
\includegraphics[scale=0.45]{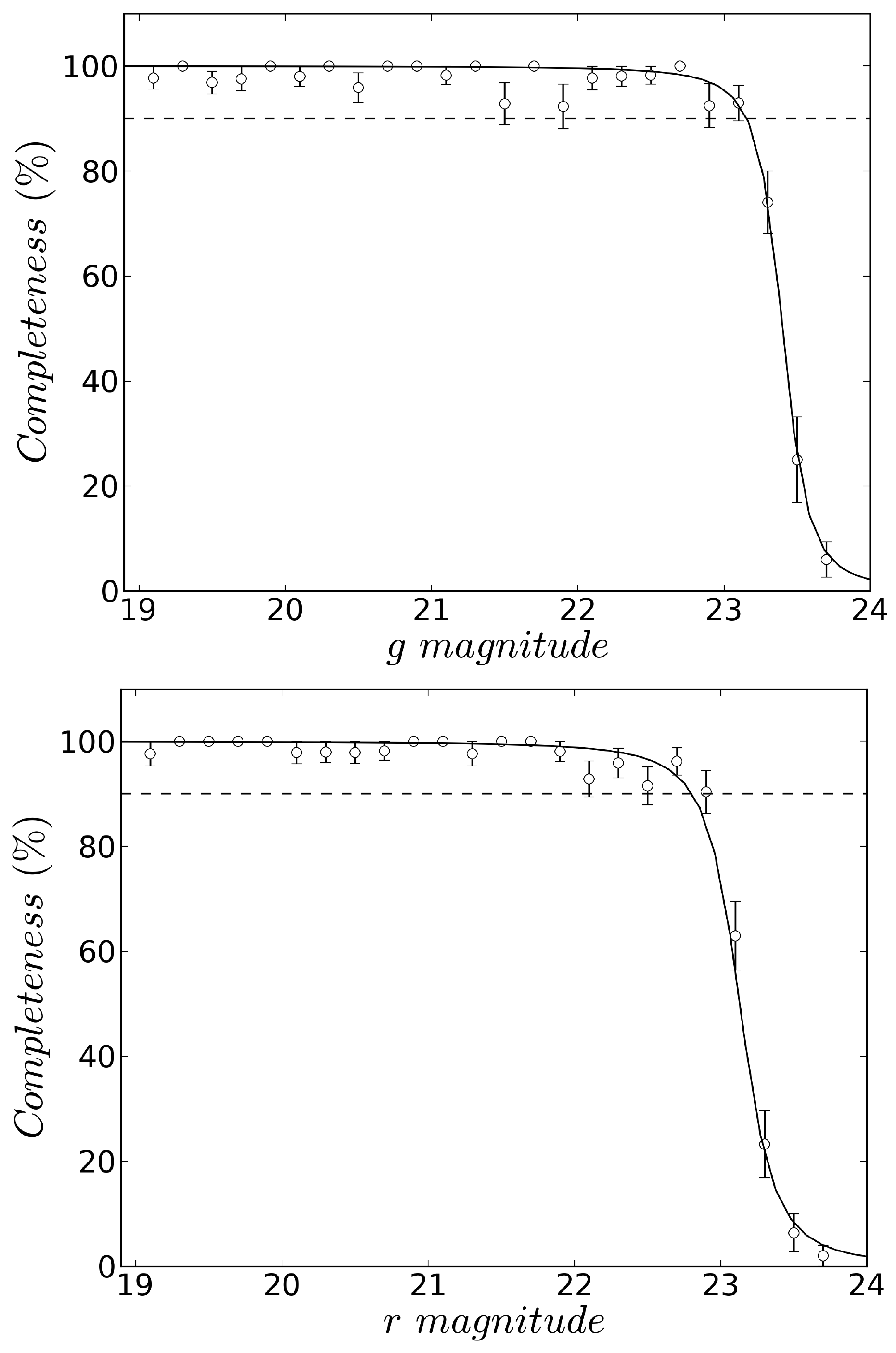}
\caption{Artificial star tests: detection completeness as function of 
magnitude in the $g$-band (\emph{upper panel)} and $r$-band (\emph{lower panel}).
Overplotted are the best-fitting analytic interpolation functions. The 90\% completeness 
levels (dashed line) correspond to $g=23.15$ and $r=22.80$ respectively.}

\label{fig:CompletenessTest}
\end{figure}

We performed artificial star tests to determine the photometric completeness as a function of magnitude. We use the {\em starlist} function in the IRAF package {\em artdata} to generate 1000 randomly distributed artificial stars for each $g$ and $r$ band in the range of $19<g,r<24$, and then employ the {\em mkobjects} function to inject them into the DECam images. We then extract the stars from the images using the same pipeline as we used on the original science images, and cross-match all the point sources with the list of input artifical stars using STILTS to measure the recovery rate as a function of magnitude. Figure~\ref{fig:CompletenessTest} shows the completeness levels for both photometric bands. The error bars for each bin are derived from the binomial distribution as in~\cite{Bolte1989}. Overplotted are the best-fitting interpolation functions described in~\cite{Fleming1995}. We determine the $90\%$ completeness levels of our photometry at $g=23.15$ and $r=22.80$, respectively. 

\begin{figure*}[t!]
\plotone{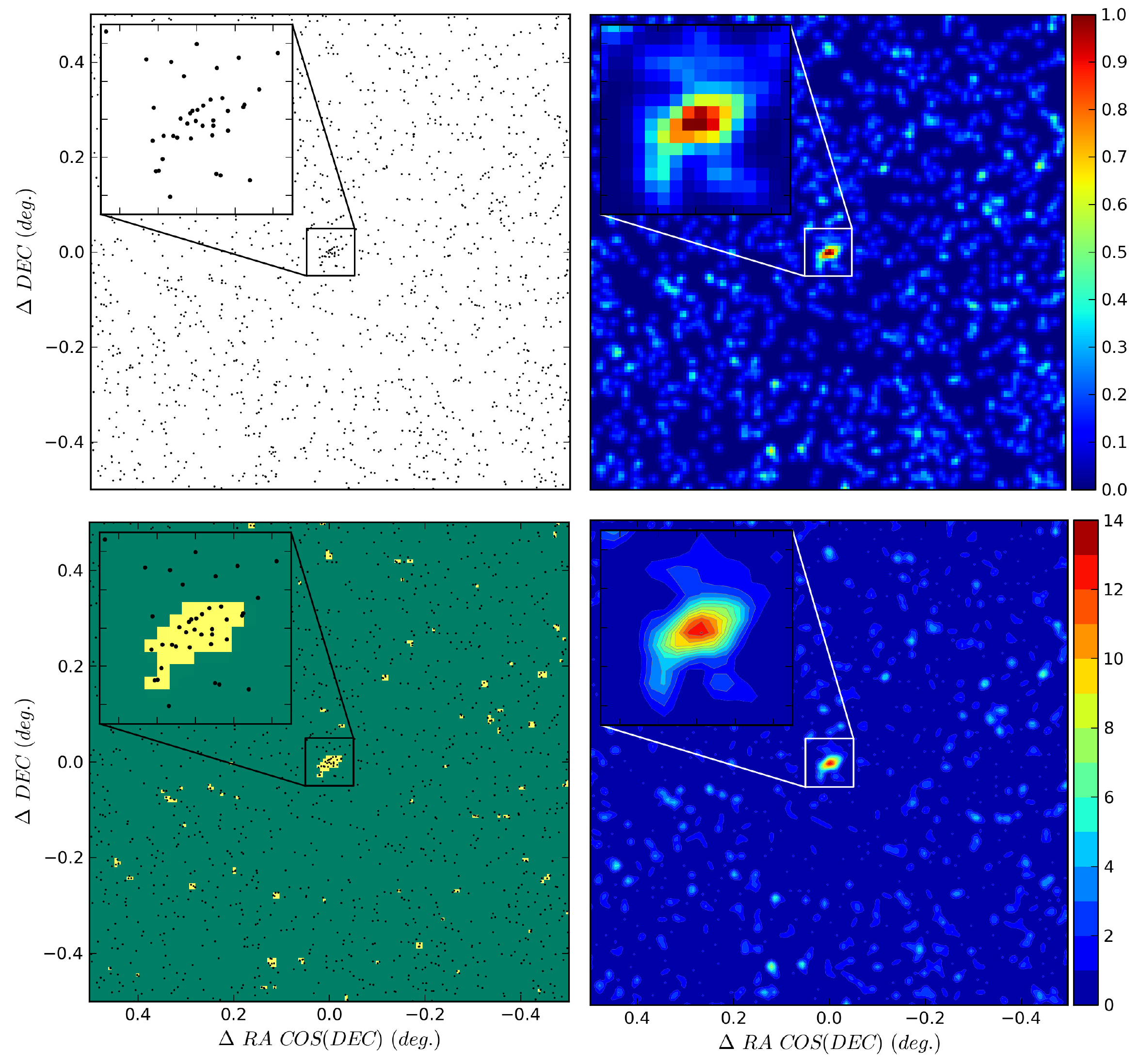}
\caption{\emph{Upper left panel}: Spatial positions of DECam stars, in the magnitude range from the saturation level to the $90\%$ completeness level for each $g$ and $r$ band, passing the photometric selection criteria. \emph{Upper right panel}: Binned spatial density with a pixel size of $23.4\,\arcsec$ and convolved with a Gaussian kernel with FWHM of $55.1\,\arcsec$. The color of the pixels represent the output signals ($s$) of the convolution. \emph{Lower left panel}: Selected bins of which the output signals are above the foreground density $\tilde{s}+2.5\sigma_{s}\approx0.3$ . \emph{Lower right panel}: Contour plot of the convolved density map. The contours show the levels of the output signals in units of $\sigma_{s}$ above the Galactic foreground.}
\label{fig:Contour}
\end{figure*}

Figure~\ref{fig:Contour} shows the stellar overdensity in the running window after applying the photometric filter with a Dartmouth isochrone of 12 Gyr and 
[Fe/H]$=-1.7$~\citep{Dartmouth}. The top left panel shows the sky positions of DECam stars fainter than the saturation limit ($r_0\approx 19$ mag) and brighter than the $90\%$ completeness level ($r_0=22.80$), passing the isochrone filtering process at the distance modulus $(m-M)=16.5$ magnitudes. The top right panel shows the spatial density plot with a pixel size $23.4\,\arcsec$ and convolved with a Gaussian kernel with a FWHM of $55.1\,\arcsec$, where the colours of the pixels represent the output signals\,($s$) of the convolution. The pixels above the foreground density with values $\tilde{s}+2.5\sigma_{s}\approx0.3$ are marked in yellow in the bottom left panel. These contiguous pixels are concentrated on the central region and there is no other comparable overdensity, in terms of both integrated signal and area, in the vicinity of Kim\,1 at the same distance. The bottom right panel shows the corresponding contour plot, where the contours represent the different levels of normalised star density in units of standard deviation  of the output signals above the Galactic foreground.

\section{Candidate Properties}
\subsection{Color-Magnitude Diagram}

\begin{figure*}[t!]
\plotone{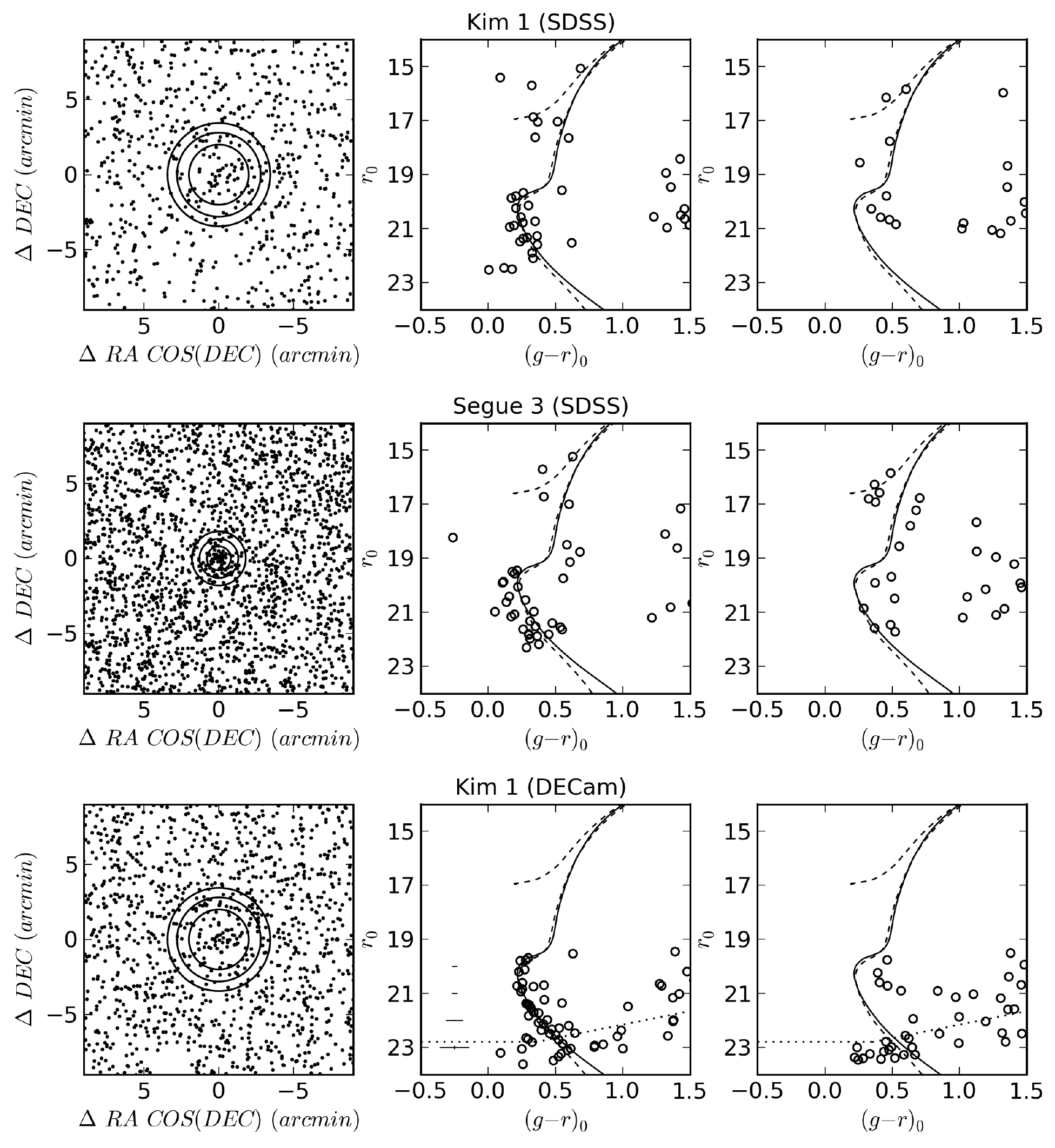}
\caption{SDSS view of Kim\,1 (top) and Segue\,3 (middle), and DECam view of Kim\,1 (bottom). \emph{Left panels}: Distribution of  all objects classified as stars in a $9\,\arcmin\times9\,\arcmin$ field centred on the star cluster. The circles mark a radius of $2.0\arcmin$, $5.0\arcmin$ and $5.4\arcmin$ for Kim\,1, and $0.8\arcmin$, $2.4\arcmin$ and $2.6\arcmin$ for Segue\,3. \emph{Middle panels}: CMD of all stars within the inner-most circle marked on the left panel, dominated by the members of the star cluster. \emph{Right panels}: Comparison CMD of all stars between the middle and the outer circles, showing the foreground stars. The dotted lines in the bottom row mark the 90\% completeness level of our photometry. The Dartmouth (solid) and \textsc{Parsec} (dashed) isochrones of age 12 Gyr and [Fe/H]=-1.7 are overplotted at a distance of 19.8 kpc for Kim\,1, and the same isochrones at $17$ kpc for Segue\,3~\citep{Fadely2011}.\label{fig:HessDiagram}}
\end{figure*}

The upper left panel of Figure~\ref{fig:HessDiagram} shows the distribution of all objects classified as stars by the SDSS pipeline in the vicinity of Kim\,1. For comparison we show the SDSS data for Segue\,3~\citep{Belokurov2010,Fadely2011} in the second row. Kim\,1 has a lower star density and appears less prominent, 
until the photometric cut has been applied. The panels in the middle column of Figure~\ref{fig:HessDiagram} show the extinction-corrected CMDs of the two overdensities, and the panels in the right column that of the foreground stars around them. The CMDs of the foreground stars have been established following the same procedure as in \citet{Belokurov2006,Belokurov2007} from the area between two concentric rings covering the same area as Kim\,1 and Segue\,3 respectively. Kim\,1 has an equally well defined main sequence (MS) as Segue\,3 down to $r_0=22$\,mag. There are also five stars with $r_{0}$ magnitudes brighter than 18\,mag and colours consistent with RGB and red clump stars. 
Finally, the panels in the bottom row of Figure~\ref{fig:HessDiagram}  show the results for Kim\,1 based on our DECam photometry reaching $\sim$2 mag fainter than SDSS with S/N values for point sources of 35 and 20, in $g$ and $r$ respectively, at 23$^{rd}$ AB mag.
The DECam data reveals a tight main sequence between $19.5<r_{0}<24.0$ mag.

\subsection{Age and Metallicity}

\begin{figure}
\includegraphics[scale=0.46]{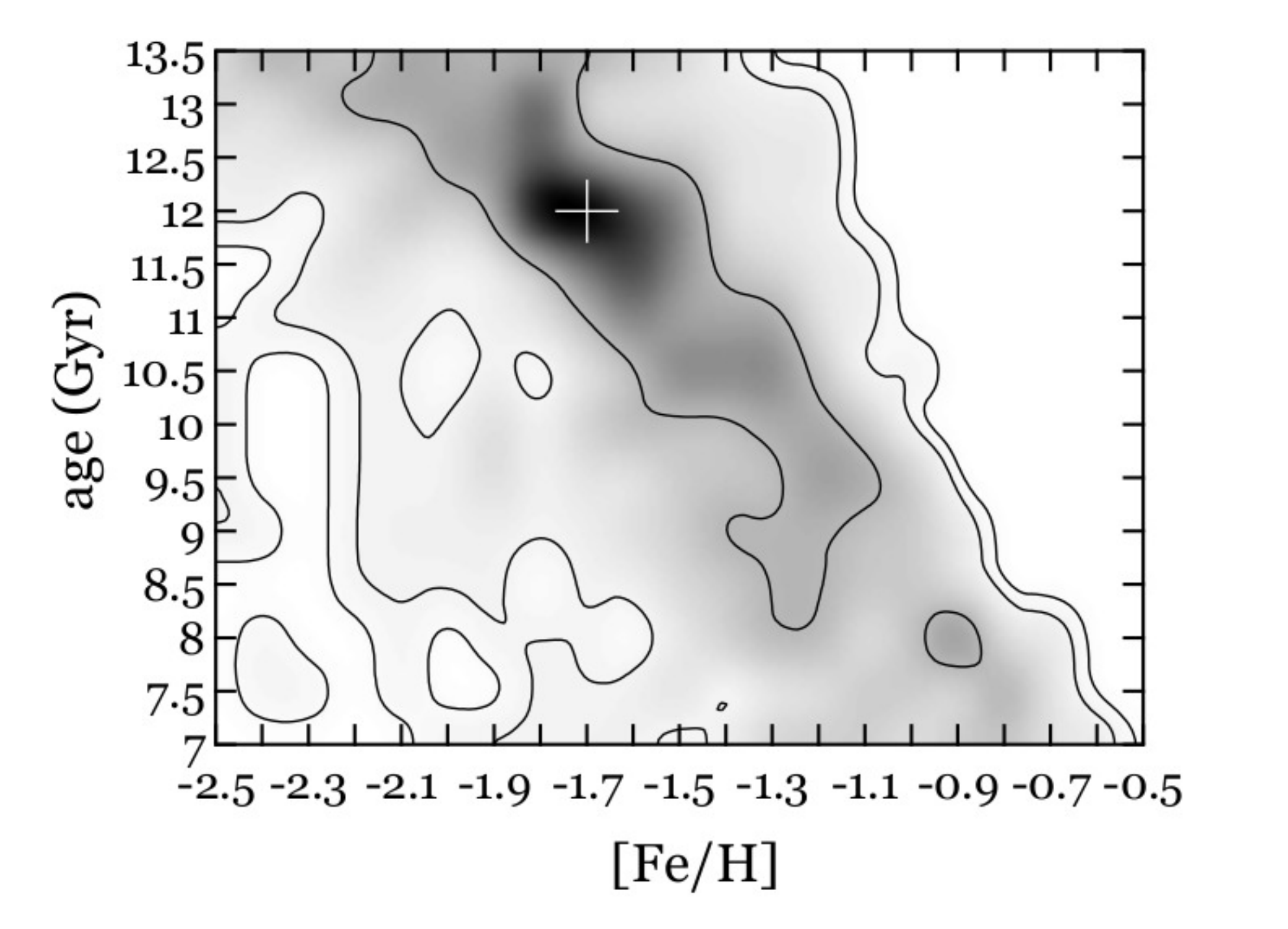}
\caption{
Smoothed maximum likelihood density map in age-metallicity space for all stars within a radius 
of $2\arcmin$ around Kim\,1, combined with SDSS ($r<19$) and DECam ($r>19$) coverage.
Contour lines show the 68\%, 95\%, and 99\% confidence levels. The diagonal flow of the contour lines reflect the
age-metallicity degeneracy inherent to such an isochrone fitting procedure. The 1D marginalized parameters 
around the best fit are:  age $=12.0^{+1.5}_{-3.0}$\,Gyr, [Fe/H]$=-1.7^{+0.6}_{-0.2}$\,dex, $m-M=16.48\pm0.10$\,mag.
}
\label{fig:MLplot}
\end{figure}

We estimate the age, [Fe/H], and distance of Kim\,1 using the maximum likelihood method described in~\cite{Frayn2002} and \cite{Fadely2011}. We use all stars within a radius of $2\arcmin$ around Kim\,1, combining SDSS ($r<19$)  and 
DECam ($r>19$) photometry. We calculate the maximum likelihood values $\mathcal{L}_i$ over a grid of Dartmouth model 
isochrones \citep{Dartmouth} covering an age range from 7.0--13.5\,Gyr, a metallicity range $-2.5\leq$ [Fe/H] $\leq-0.5$\,dex, and a distance range 
$16<(m-M)<17$.  Grid steps are 0.5\,Gyr, 0.1\,dex, and 0.05\,mag.  In Figure\,\ref{fig:MLplot}, we present the matrix of likelihood values for the sample described above
after interpolation and smoothing over 2 grid points.  We find the Dartmouth isochrone with the 
highest probability has an age of 12.0 Gyr and [Fe/H] $= -1.7$\,dex. The 68\%, 95\%, and 99\% confidence contours are also overplotted in 
the figure.  The marginalized uncertainties about this most probable location correspond to an age of 
$12.0^{+1.5}_{-3.0}$\,Gyr, a metallicity of [Fe/H]$=-1.7^{+0.5}_{-0.2}$\,dex, and a distance modulus of 
$\rm (m-M)_0=16.48\pm0.10$\,mag ($d_\odot = 19.8\pm0.9$ kpc).  We adopt a heliocentric distance of 19.8\,kpc in the calculation of 
physical size and absolute magnitude in Section 4.2.

The best-fitting Dartmouth~and \textsc{Parsec}~\citep{PARSEC} isochrones of age 12 Gyr and [Fe/H]$=-1.7$ that 
match the location of the RGB, red clump and MS of Kim\,1 are overplotted at a distance of $19.8$ kpc to illustrate 
the consistency with an old metal-poor stellar population independently of the theoretical model. Segue\,3 happens to have the same age and metallicity ~\citep{Fadely2011}. We adopted a distance of 17\,kpc for 
the isochrone shown in the CMD of Segue\,3.

\subsection{Size, Ellipticity, and Luminosity}

\begin{figure}
\plotone{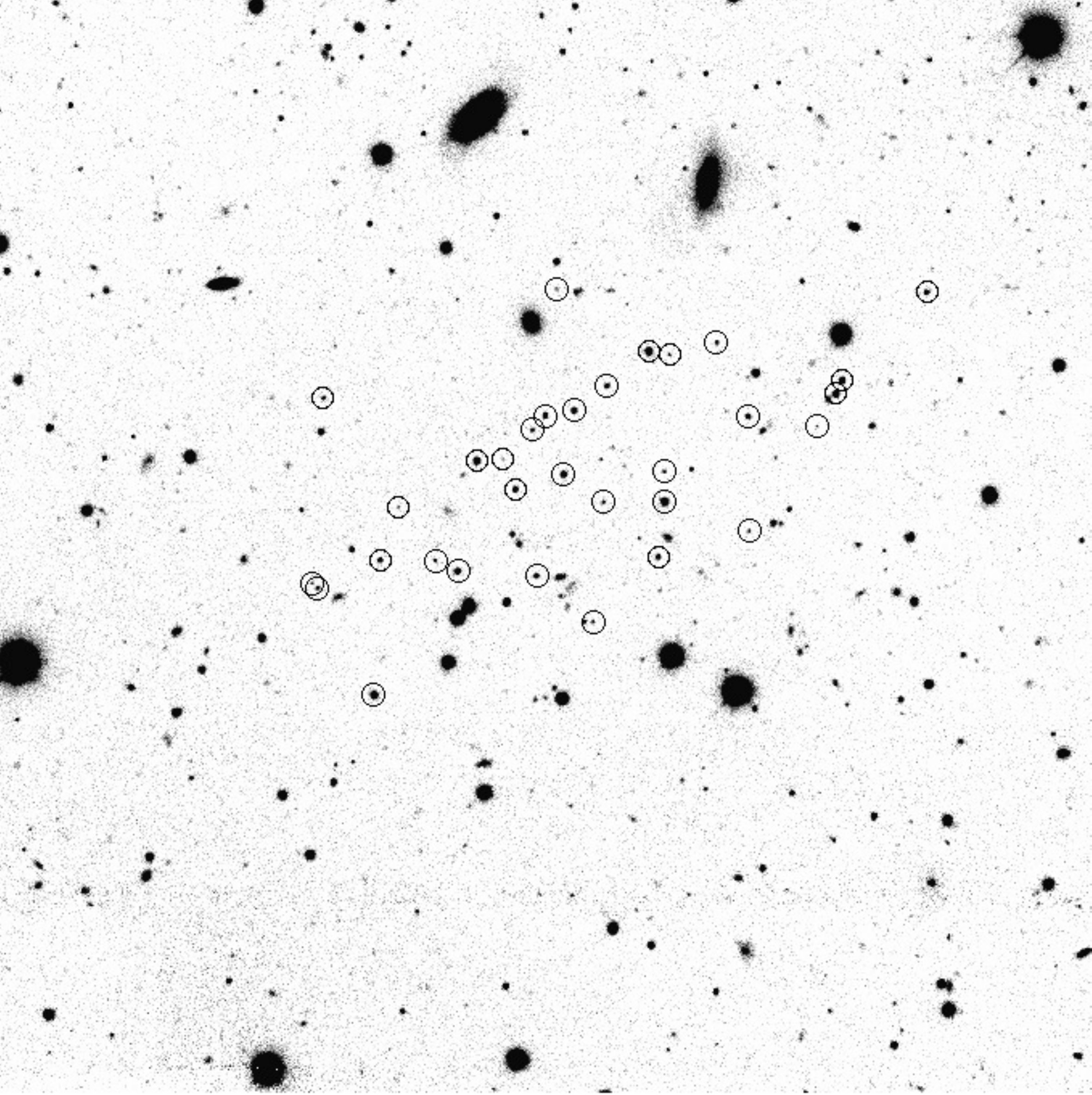}
\caption{$6\times 6$\,arcmin$^2$ DECam cutout g-band image of Kim\,1. North is up, east is to the left. Circled are all stars within $2r_{h}$ from Kim\,1 that are
consistent in color-magnitude space with the best-fitting isochrone (12 Gyr, [Fe/H]=-1.7) at a distance of 19.8\,kpc. The main-sequence stars are fainter than $r_0\approx 19.5$ and have colours within $0.1<(g-r)_0<0.6$.\label{fig:Fits-g}}
\end{figure}

Figure~\ref{fig:Fits-g} shows the $6\times 6$\,arcmin$^2$ DECam cutout $g$ band image of Kim\,1. The circles indicate the locations of all stars within  $2r_{h}$ from Kim\,1 in Figure~\ref{fig:HessDiagram} that passed our photometric cut based on the isochrone-fit in color-magnitude space.

\begin{figure*}[t!]

\begin{centering}
\includegraphics[scale=0.85]{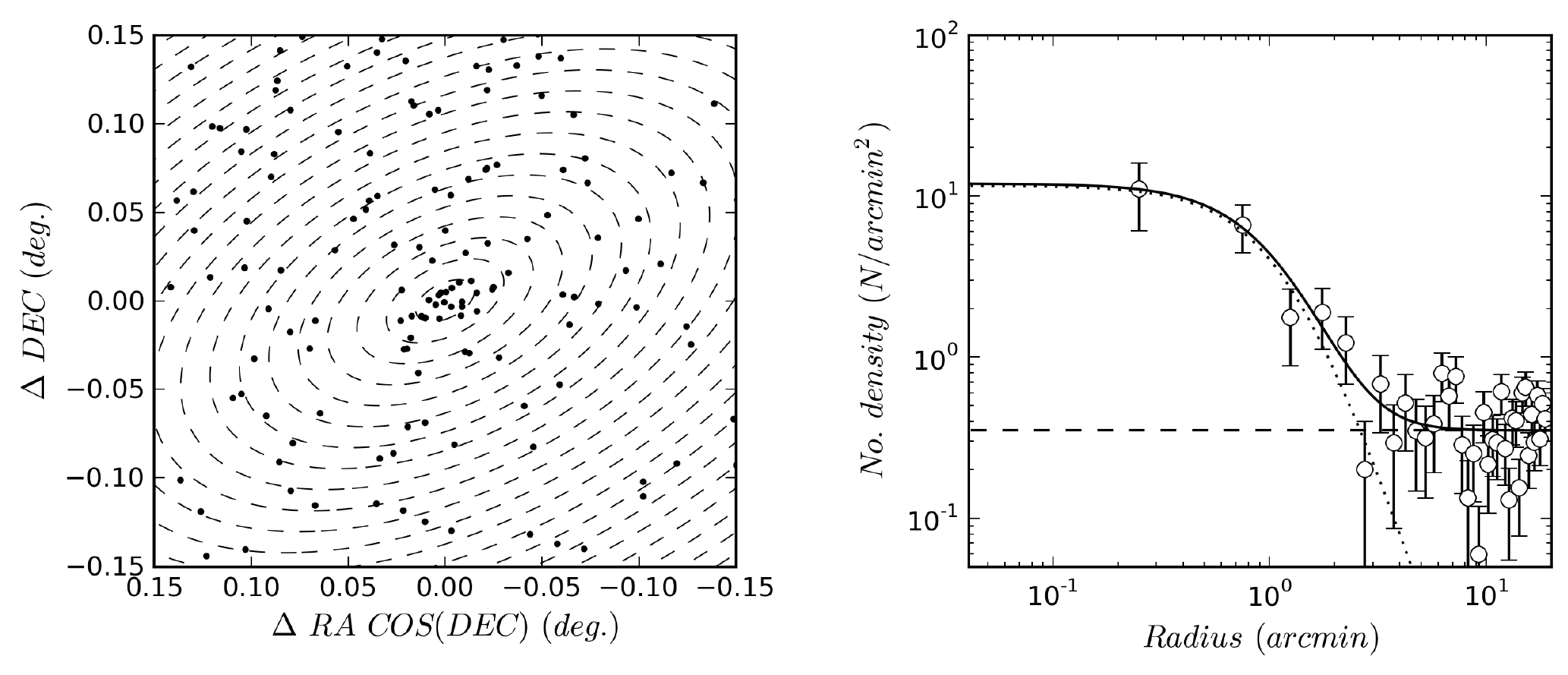}
\end{centering}
\protect\caption{\emph{Left panel}: RA-DEC distribution of DECam stars passing the photometric selection criteria centred on Kim\,1. The dashed ellipses indicate $1.5\,\arcmin$ steps in the elliptical radius of ellipcity $\epsilon=0.42$ and position angle $\theta=-59\,\deg$. \emph{Right panel}: Radial stellar density profile based on the stellar distribution in the left panel. The dotted line shows a Plummer profile with a scale parameter $a=1.2\arcmin$ The dashed line shows the contribution of foreground stars. The solid line is the combined fit.\label{fig:Plummer}}
\end{figure*}

The left panel of Figure~\ref{fig:Plummer} shows the sky distribution of all DECam stars, in the magnitude range from the saturation level to the $90\%$ completeness level for each $g$ and $r$ band, passing the photometric cut 
based on the best-fitting model isochrone, in a $0.3^\circ\times0.3^\circ$ window centred on Kim\,1. The right panel shows the associated radial profile of the stellar number density, i.e. star counts in elliptical annuli around Kim\,1, 
where $r$ is the elliptical radius. 
For the center of the overdensity we adopted the  centre of mass calculated in terms of the normalised signals on the array into which the stars were binned and smoothed as described in Section 2. We derive an ellipticity $\epsilon=0.42\pm0.10$ and a position angle $\theta=-59\pm6$ deg by using the fit\_bivariate\_normal function of the astroML package~\citep{astroML}. The error bars were derived from Poisson statistics. Overplotted is the best-fit Plummer profile \citep{PlummerModel} to parametrise the underlying stellar distribution. We obtain a half-light radius of $1.2\pm0.1$\,arcmin or $r_h=6.9\pm0.6$\,pc, adopting the distance modulus of $16.48$\,mag.

In analogy to \cite{Walsh2008} we estimate the total luminosity of Kim\,1 by integrating the radial number density profile shown in Figure~\ref{fig:Plummer} to calculate the total number of Kim\,1 stars $N_{*}$ within the photometric limits. Using the ratio of this number $N_{*}$ to the probability density of a normalised theoretical luminosity function (LF) in the same magnitude range as for the radial profile, we scale the theoretical LF to the number densities of stars as a function of $r$ magnitude. Integrating the scaled LF inclusive of the missing flux from stars brighter than the saturation limit and fainter than the $90\%$ completeness limit, gives a total luminosity $M_{r, Kim\,1}=0.04$\,mag based on the PARSEC isochrone and $M_{r, Kim\,1}=-0.07$\,mag based on the Dartmouth isochrone, both for a 12 Gyr old stellar population 
with [Fe/H]$=-1.7$. Due to the intrinsic faintness of the stellar overdensity and the low number of stars, the estimates of the total luminosity has 
a large uncertainty. From the $V-r$ colours ($0.32$\,mag and $0.22$\,mag) of the two best-fitting model isochrones we derive the corresponding $V$-band luminosities $M_{V, Kim\,1}=0.36$\,mag and $M_{V, Kim\,1}=0.15$\,mag, respectively. We note that both model isochrones give consistent results. Since this method relies on the number of stars in such an ultra-faint stellar system instead of their individual flux, the inclusion or exclusion of a single RGB or red clump star in the system can change its total luminosity still by $\sim0.5$ mag. Hence, a realistic estimate of the total luminosity of Kim\,1 is $M_{V}=0.3\pm0.5$. All derived parameters are summarised in Table~\ref{tab:Parameters}. 

\begin{deluxetable}{lrl}
\tablewidth{0pt}
\tablecaption{Properties of Kim\,1}
\tablehead{
\colhead{Parameter} & 
\colhead{Value} &
\colhead{Unit}}
\startdata
$\alpha_{J2000}$ & 22:11:41.3 & h:m:s \\
$\delta_{J2000}$ & +07:01:31.8 & $^\circ:\arcmin:\arcsec$ \\
$l$ & 68.5148 & deg\\
$b$ & -38.4256 & deg\\
$(m-M)$ & $16.48\pm0.10$ & mag \\
$d_\odot$ & $19.8\pm0.9$ & kpc \\
{[Fe/H]} & $-1.7^{+0.5}_{-0.2}$ & dex \\
Age & $12^{+1.5}_{-3.0}$ & Gyr \\
$r_{h}$(Plummer) & $6.9\pm0.6$ \tablenotemark{a} & pc \\
$\epsilon$ & $0.42\pm0.10$ & \\
$\theta$ & $-59\pm6$ & deg \\
$M_{tot,V}$ & $0.3\pm0.5$ & mag \\
\enddata
\tablenotetext{a}{ Adopting a distance of 19.8\,kpc}
\label{tab:Parameters}
\end{deluxetable}

\section{Discussion and Conclusion}
We report the new ultra-faint stellar overdensity, Kim\,1, in the constellation of Pegasus. The best-fitting theoretical isochrone reveals a single 12\,Gyr old, metal-poor ([Fe/H]=-1.7) stellar population at a well-defined heliocentric distance of $19.8\pm0.9\,$kpc. Its total luminosity of $M_{V}=0.3\pm0.5$ shows that Kim\,1 is among the faintest stellar systems discovered in the Milky Way halo to date. Other MW star clusters known to have comparable luminosities are Segue\,3~\citep[$M_{V}=0.0\pm0.8$;][]{Fadely2011}, Mu\~noz \,1~\citep[$M_{V}=-0.4\pm0.9$;][]{Munoz2012}, Balbinot\,1~\citep[$M_{V}=-1.21\pm0.66$;][]{Balbinot2013}, and Koposov\,1 \& 2~\citep[$M_{V}=-2$, $M_{V}=-1$;][]{Koposov2007}. Its physical size ($r_h=6.9\pm0.6$\,pc) and low luminosity place Kim\,1 close to these MW star clusters in the size-luminosity diagram~\citep{Fadely2011}.

\subsection{Kim\,1, a dissolving star cluster?}
Despite the fact that the stellar population of Kim\,1 appears sufficiently old to be dynamically relaxed for the given physical size and the number of constituent stars, the overdensity has an unusually high ellipticity of 0.42 and a poorly defined central concentration. No other low luminous MW star cluster, with the exception of Koposov 1 \citep[][see their Figure 2a]{Koposov2007} has comparable structural properties, which suggest that we are seeing a tidally 
disrupted star cluster or a remnant thereof. We note that the small size of Kim\,1 and the absence of a more complex stellar population make it unlikely 
that Kim\,1 is a ultra-faint satellite galaxy of the Milky Way~\citep{Willman2012}. 

\begin{figure*}
\begin{centering}
\includegraphics[scale=0.5]{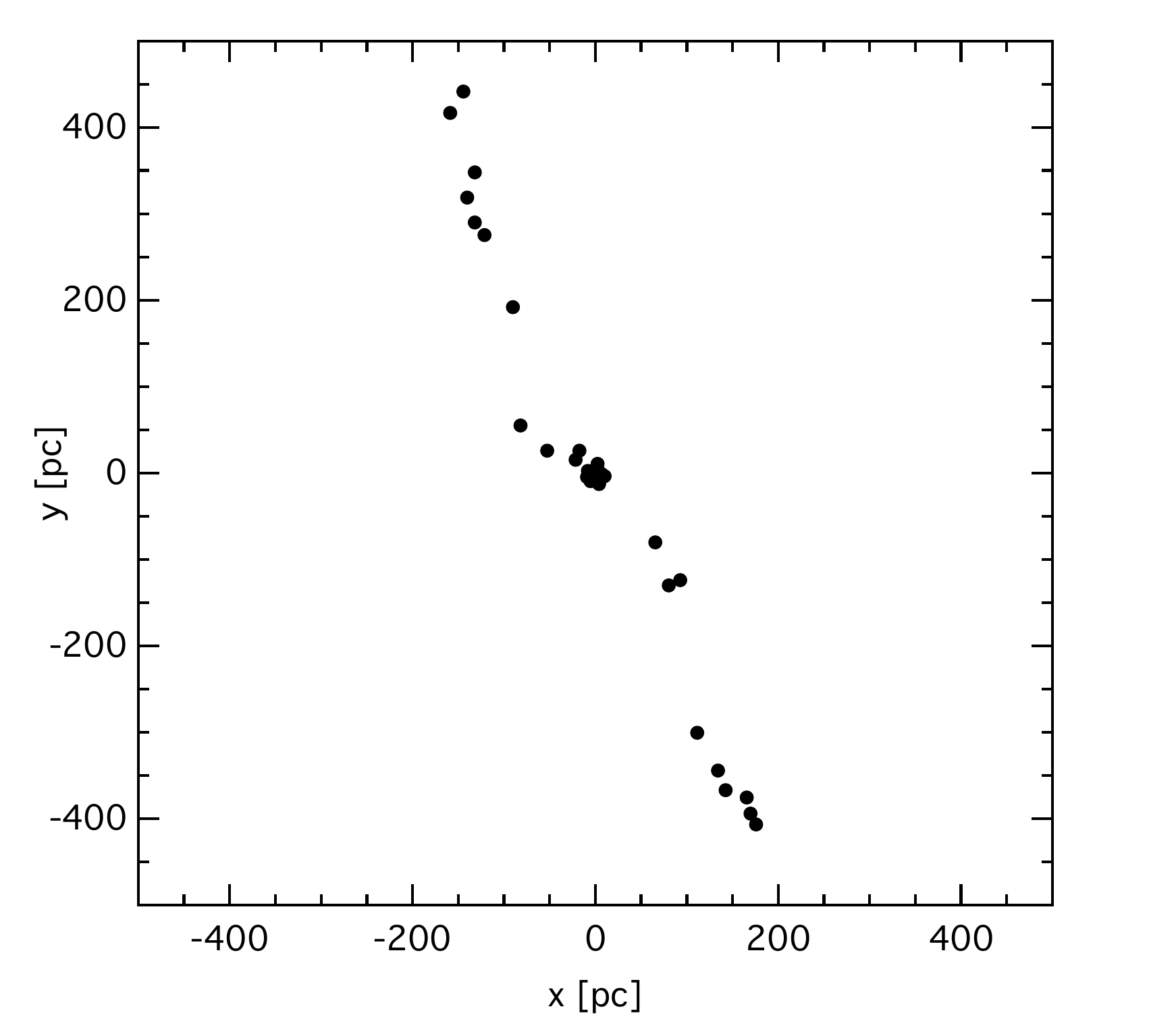}
\includegraphics[scale=0.5]{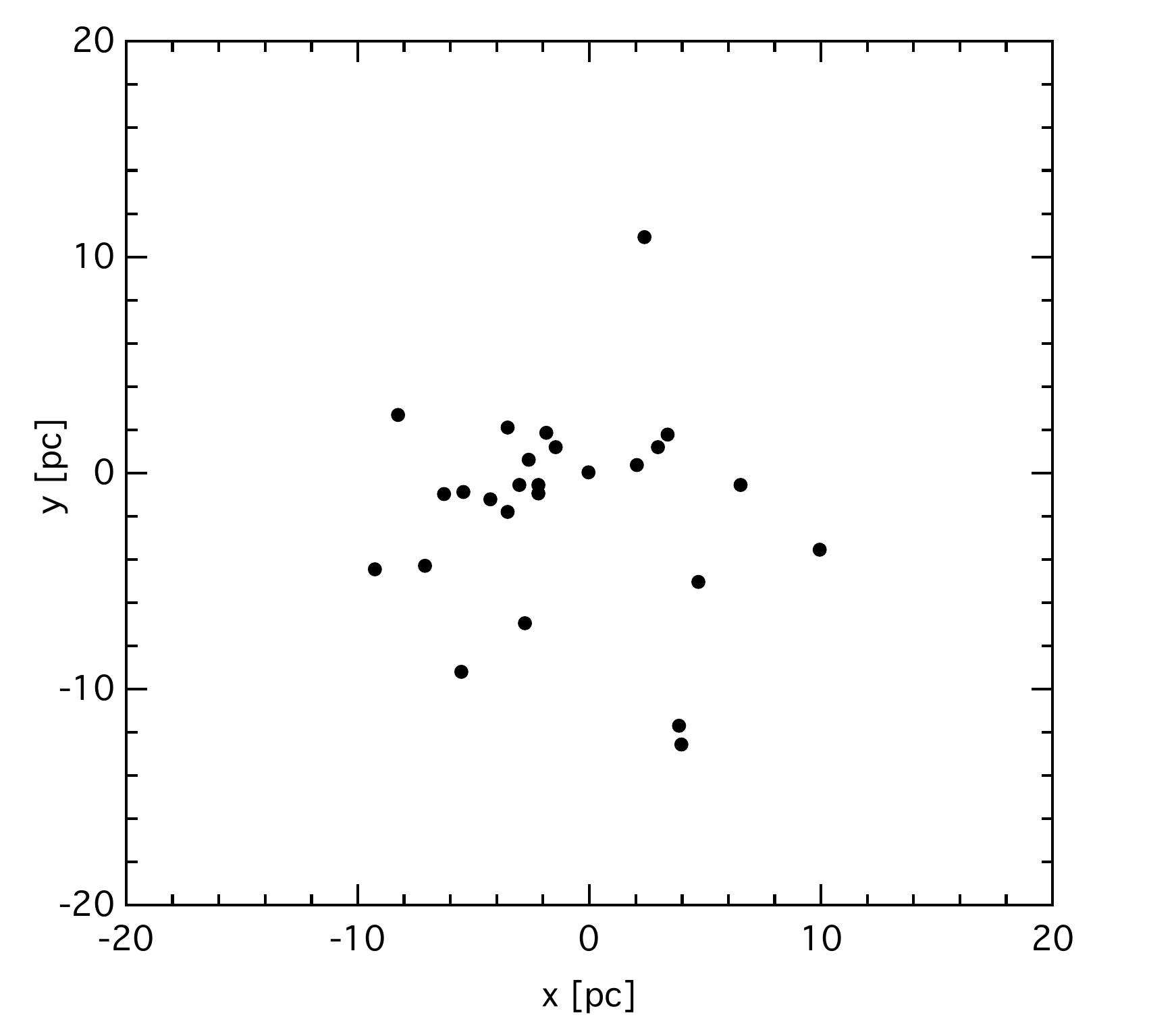}
\par\end{centering}
\protect\caption{Result of N-body simulation assuming a spherical Milky Way potential with a constant rotation velocity of 
$V_G=220$\,km\,s$^{-1}$ and a model star cluster with $M_{ini}=5500 M_{\odot}$ in a circular orbit at the distance of 25 kpc. \emph{Left panel}: Line-of-sight stellar 
distribution of the model star cluster after 12\,Gyr for the DECam field-of-view. \emph{Right panel}: The same as the left panel but for the inner region.
Star counts and stellar distribution closely resemble Kim\,1.}
\label{fig:Nbody}
\end{figure*}

Although a detailed simulation of Kim\,1's dynamical history 
is beyond the scope of this study due to the lack of kinematical data 
and orbital parameters, Baumgardt (private communication) kindly ran 
N-body simulations using NBODY6~\citep{Aarseth1999} to estimate the initial mass of the progenitor star cluster,
using a set-up similar to ~\cite{Baumgardt2003}. 
For that purpose a model star cluster with a~\cite{Kroupa2001} initial mass function was 
placed on circular or eccentric orbits in a spherical Milky Way potential with a constant rotation velocity of 
$V_G=220$\,km\,s$^{-1}$. The integration of the orbits was halted after 12\,Gyr, 
the age of Kim\,1. Matching the bound mass of the remnant model star cluster with the current mass 
of Kim\,1 (approximately $100\,M_\odot$) confines its initial stellar mass to the range $3500 M_{\odot} < M_{init}<5500 M_{\odot}$, 
where the two limits correspond to a highly eccentric orbit with a perigalactic and apogalactic distance of 20\,kpc and 100\,kpc respectively,
and a circular orbit at the distance of 25\,kpc. In Figure~\ref{fig:Nbody} we show the line-of-sight stellar 
distribution of the model star cluster in the circular orbit after 12\,Gyr for the DECam field-of-view (left) 
as well as the inner region (right). There
are only a small number of extra tidal stars distributed over an area corresponding to the DECam field, 
and the remnant star cluster is highly asymmetric and elongated in the inner $30$\,pc$\times30$\,pc  without a well-defined center. 
From these simulations we can conclude that the observational properties of Kim\,1 are consistent with a 
dissolving star cluster that initially had a few thousand solar masses. No prominent stellar stream is expected to be associated 
to such a low mass star cluster. The few cluster stars found beyond the tidal radius would be completely hidden in the screen of stars from the Milky Way.

\subsection{Is Kim\,1 associated to a Stream?}
The constellation of Pegasus, in which Kim\,1, Segue\,3 and Balbinot\,1 are found, lies close to the Hercules-Aquila Cloud discovered in SDSS-DR5~\citep{HERCULES-AQUILA CLOUD}. This Galactic halo substructure is centred at $l\sim40^\circ$, and extends down to $b\sim-40^\circ$ at $l\sim50^\circ$. In fact, \citet{deJong2010} used main-sequence stars in SDSS-DR7 SEGUE data to identify a number of new halo overdensities that might be related to the cloud. As Segue\,3 coincides spatially with one of the reported overdensities, there has been speculations about the possible association of Segue\,3 with the Hercules-Aquila Cloud. A spectroscopic analysis~\citep{Fadely2011}, however, found a kinematic offset that suggests Segue\,3 is unlikely associated with that overdensity. For Balbinot 1 and Kim\,1 located at $(l,b) = (75.1735^\circ, -32.6432^\circ)$ and $(l,b) = (68.5148^\circ$, $-34.4256^\circ$) respectively, there is no corresponding overdensity out of all the detections listed in Table 4 of \cite{deJong2010}. In addition, there is an argument that the true heliocentric distance of the Hercules-Aquila Cloud ranges only between 1 and 6 kpc~\citep{Larsen2011}. \cite{Simion2014} conclude in their recent study using RR Lyrae stars in the Catalina Sky Survey data that the extension of the  Hercules-Aquila Cloud is bound within $30^\circ < l < 55^\circ$ and $-45^\circ < b < -25^\circ$, and that there is no significant excess of stars in the region of sky at $55^\circ < l < 85^\circ$ and $-45^\circ < b < -25^\circ$. Since Kim\,1, Segue\,3, and Balbinot 1 are all located at $ l > 68^\circ $, these star clusters are unlikely associated with the Hercules-Aquila Cloud. 

\begin{figure*}

\begin{center}
\includegraphics[scale=0.8]{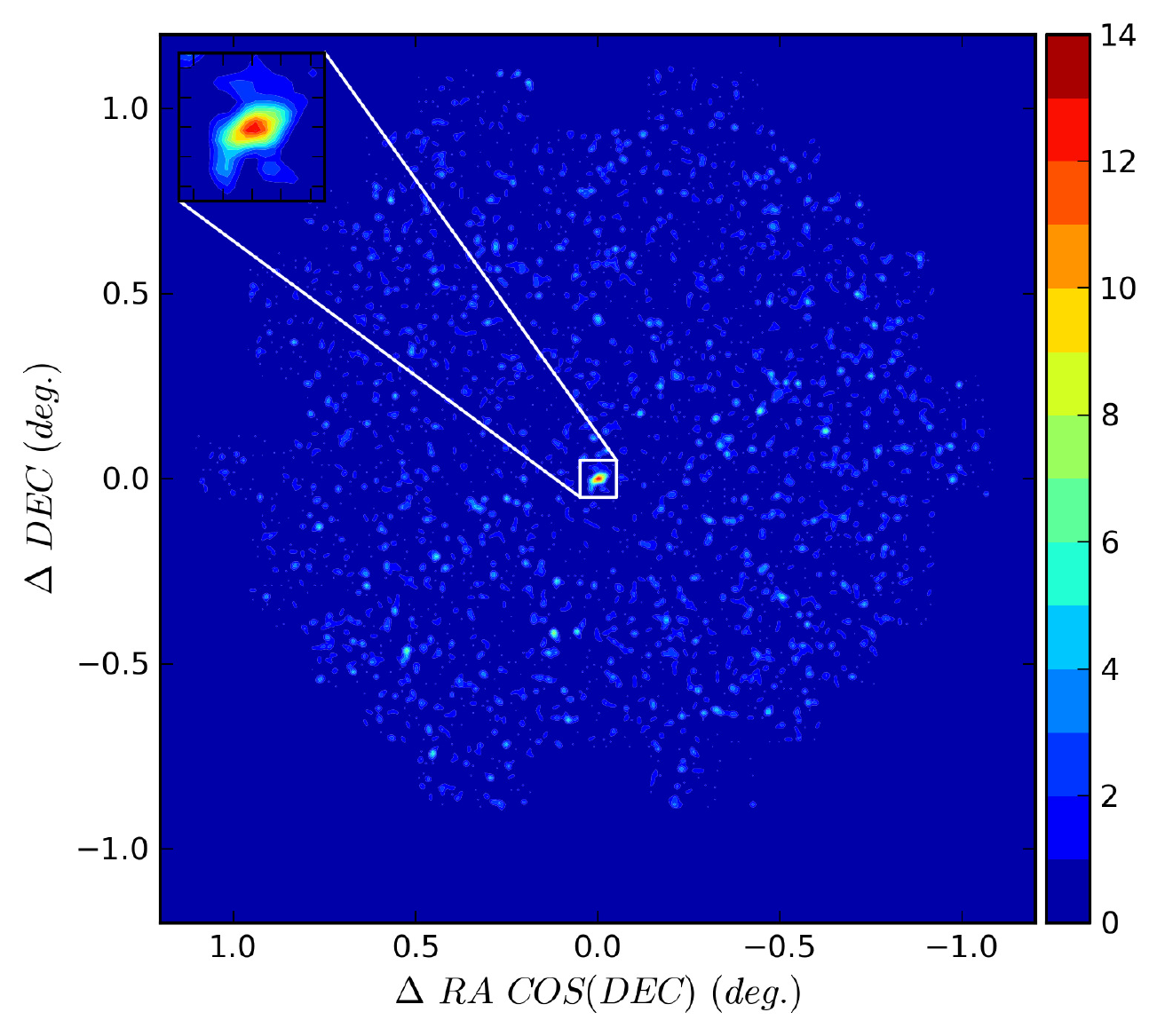}
\end{center}
\caption{The same as the lower right panel of Figure~\ref{fig:Contour} but for all stars in the field of view of DECam. }
\label{fig:DECamView}
\end{figure*}

Recent searches of SDSS DR8 and DR10 for extended stellar overdensities in the Galactic halo have discovered new stellar streams in a range of heliocentric distances from 15 to 35 kpc~\citep{Bonaca2012,Martin2013,Grillmair2014}. These stellar streams are identified on extended stellar density maps by taking photometric cuts with weighted isochrone masks over a range of distance moduli. Due to the sparse and extended nature of the streams, their stellar populations are barely discerned from the Milky Way foreground in color-magnitude space unless a field-subtracted Hess diagram is constructed out of a sufficiently large sky area. None of those searches, however, have found any stellar stream that overlaps with the location of Kim\,1~\citep[e.g. Figure 2 in][]{Bonaca2012}, and the stellar population of Kim\,1 is clearly distinguished from the foreground in the CMD of a localised region. In addition, unlike tidal debris detected around Pal 5~\citep{Odenkirchen2001}, our DECam star distribution does not feature any extra significant clumps around Kim\,1 out to 400\,pc radius. There is also a hypothesis that ultra-faint dwarf galaxies are instead cusp caustics of cold stellar debris that formed during the disruption of low-mass satellites~\citep{Zolotov2011}. A cusp caustic in projection appears as a highly elongated center of stars, along with a two-fold of tails as an indication. Although Kim\,1 indeed appears highly elongated compared to other known ultra-faint star clusters, such small scale tails and their fold are not detected in our DECam data (see Figure~\ref{fig:DECamView}). 

It is worth noticing that Segue\,3 ($d_\odot=17$\,kpc, 12 Gyr, [Fe/H]=-1.7), Kim\,1 ($d_\odot=19.8$\,kpc, 12 Gyr, [Fe/H]=-1.7), and Balbinot 1 ($d_\odot=31.9$\,kpc, 11.7 Gyr, [Fe/H]=-1.58), are neighbours in projection and even lie within a considerably narrow range of heliocentric distances from 17 to 32\,kpc. They are also spatially close to the globular cluster M2~\citep[$d_\odot$=11.5\,kpc;][]{Harris1996}, one of the most unusual globular clusters in the Milky Way halo with a number of distinct stellar subpopulations (Milone et al. 2014, submitted), and their locations coincide with the vast polar structure (VPOS), a thin (20\,kpc) plane perpendicular to the MW disk defined by the 11 brightest Milky Way satellite galaxies~\citep{Kroupa2005,Metz2007,Metz2009,Kroupa2010,Pawlowski2012}. Globular clusters and stellar and gaseous streams appear to preferentially aligned with the VPOS too~\citep{Forbes2009,Pawlowski2012}. It might be a coincidence that all four stellar systems mentioned above are spatially close together and that their locations agree with the VPOS, but this finding naturally raises the question about these star clusters' possible link to the Milky Way satellite galaxies. While more information is required to address Kim\,1's origin, the finding of such a small, extreme low luminosity stellar concentration in the inner halo is definitely consistent with current detection limits (Walsh et al. 2009, see Fig.\,15) and points to a much larger parent population of star clusters that are still undetected in the Milky Way halo or must have been already destroyed by the Galactic tidal field.

The authors like to thank Tammy Roderick, Kathy Vivas and David James for their assistance during the DECam observing run. 
We gratefully acknowledge Holger Baumgardt for producing the N-body results described in the discussion. 
We also thank the referee for the helpful comments and suggestions, which contributed to improving 
the quality of the publication. We acknowledge the support of the Australian Research Council (ARC) through Discovery project 
DP120100475 and financial support from the Go8/Germany Joint Research Co-operation Scheme. Funding for SDSS-III has been provided by the Alfred P. Sloan Foundation, the Participating Institutions, the National Science Foundation, and the U.S. Department of Energy Office of Science. The SDSS-III web site is http://www.sdss3.org/.

SDSS-III is managed by the Astrophysical Research Consortium for the Participating Institutions of the SDSS-III Collaboration including the University of Arizona, the Brazilian Participation Group, Brookhaven National Laboratory, Carnegie Mellon University, University of Florida, the French Participation Group, the German Participation Group, Harvard University, the Instituto de Astrofisica de Canarias, the Michigan State/Notre Dame/JINA Participation Group, Johns Hopkins University, Lawrence Berkeley National Laboratory, Max Planck Institute for Astrophysics, Max Planck Institute for Extraterrestrial Physics, New Mexico State University, New York University, Ohio State University, Pennsylvania State University, University of Portsmouth, Princeton University, the Spanish Participation Group, University of Tokyo, University of Utah, Vanderbilt University, University of Virginia, University of Washington, and Yale University.

This project used data obtained with the Dark Energy Camera (DECam), which was constructed by the Dark Energy Survey (DES) collaborating institutions: Argonne National Lab, University of California Santa Cruz, University of Cambridge, Centro de Investigaciones Energeticas, Medioambientales y Tecnologicas-Madrid, University of Chicago, University College London, DES-Brazil consortium, University of Edinburgh, ETH-Zurich, University of Illinois at Urbana-Champaign, Institut de Ciencies de l'Espai, Institut de Fisica d'Altes Energies, Lawrence Berkeley National Lab, Ludwig-Maximilians Universitat, University of Michigan, National Optical Astronomy Observatory, University of Nottingham, Ohio State University, University of Pennsylvania, University of Portsmouth, SLAC National Lab, Stanford University, University of Sussex, and Texas A$\&$M
University. Funding for DES, including DECam, has been provided by the U.S. Department of Energy, National Science Foundation, Ministry of Education and Science (Spain), Science and Technology Facilities Council (UK), Higher Education Funding Council (England), National Center for Supercomputing Applications, Kavli Institute for Cosmological Physics, Financiadora de Estudos e Projetos, Fundação Carlos Chagas Filho de Amparo a Pesquisa, Conselho Nacional de Desenvolvimento Científico e Tecnológico and the Ministério da Ciência e Tecnologia (Brazil), the German Research Foundation-sponsored cluster of excellence "Origin and Structure of the Universe" and the DES collaborating institutions.


\end{document}